\documentclass[prd, twocolumn, nofootinbib, floatfix]{revtex4-2}
\usepackage{amsmath}
\usepackage{graphicx}
\usepackage{dcolumn}
\usepackage{bm}
\usepackage{epsfig}
\usepackage{amssymb,latexsym,mathrsfs}
\usepackage{graphicx}
\usepackage{color}
\usepackage{hyperref}

\usepackage{float}

\hypersetup{
    colorlinks=true,
    linkcolor=red,
    citecolor=blue,
}

\newcommand{\be}{\begin{equation}}
\newcommand{\ee}{\end{equation}}
\newcommand{\bs}{\begin{split}} 
\newcommand{\bea}{\begin{eqnarray}}
\newcommand{\eea}{\end{eqnarray}}

\newcommand{\om}{\Omega_m}
\newcommand{\ode}{\Omega_{\rm de}}

\newcommand{\lcdm}{$\Lambda$CDM} 
\newcommand{\fsig}{f\sigma_8}

\begin{document}

\title{The Rise of Dark Energy} 

\author{Eric V.\ Linder$^{1,2}$} 
\affiliation{${}^1$Berkeley Center for Cosmological Physics \& Berkeley Lab, 
University of California, Berkeley, CA 94720, USA\\
${}^2$Energetic Cosmos Laboratory, Nazarbayev University, 
Nur-Sultan 010000, Kazakhstan
}

%%%%%%%%%%%%%%%%%%%%%%%%%%%%%%%%%%%%%%%%%%%%%%%%%%%%%%%%%%%%%%%%%%%%%%%%
\begin{abstract} 
While dark energy has dominated cosmic dynamics only since $z\approx0.7$, 
its energy density was still $\gtrsim5\%$ of the total out to $z\approx2.5$. 
We calculate model independent constraints on its fraction from future galaxy 
surveys, finding that the rise of dark energy could be detected at $3\sigma$ 
nearly out to $z\approx2.5$. 
\end{abstract}

\date{\today} 

\maketitle

%%%%%%%%%%%%%%%%%%%%%%%%%%%%%%%
\section{Introduction}

A frontier for the next generation of cosmic surveys after the ones 
now getting underway will be pushing to higher redshift, earlier in 
the evolution of the universe. Cosmicly, this is not simply a dull 
era of complete matter domination: the dark energy density at redshift 
$z=2$ is expected to be about double the current baryon density and 
one quarter the current matter density. How dark energy rises to 
dominate the current expansion may also offer clues to its nature. 
For example, at $z=2.5$ the dark energy density fraction in different viable 
models could be 50\% greater (or less) than in the cosmological constant 
(\lcdm) case. 

We explore whether such distinctions are within the reach of next 
generation surveys of distance and structure growth. Such experiments 
being conceived include DESI2 \cite{desi2}, 
FOBOS \cite{fobos}, LSSTspec \cite{lsstspec}, 
Mauna Kea Spectroscopic Explorer \cite{mse}, 
MegaMapper \cite{mega}, SpecTel \cite{spectel}, not 
to mention those using not galaxies as tracers but intensity mapping, 
Lyman alpha forest, etc.\ (see, e.g., \cite{bull}). 
A wide variety of cosmic science cases at 
$z\gtrsim2$ is discussed in \cite{simone}. 

Section \ref{sec:models} illustrates the relevance of the 
$z\approx1.5$--2.5 redshift range for mapping the rise of dark energy. 
Section \ref{sec:method} describes the method we use to constrain the 
dark energy density at various redshifts, and Sec.~\ref{sec:results} 
presents the results. We discuss and conclude in Sec.~\ref{sec:concl}.

%%%%%%%%%%%%%%%%%%%%%%%% 
\section{Dark Energy Rising} \label{sec:models} 

Today the fractional contributions to the total energy density of the 
universe are approximately 70\% dark energy, 25\% cold dark matter, and 
5\% baryons. Their dynamical influence however also depends on their 
equation of state $w$, as $1+3w$, so dark energy with $w\approx-1$ has 
twice the impact as nonrelativistic matter with $w=0$. Thus, though 
dark energy-matter equality occurs at $z\approx0.3$, the cosmic expansion 
began accelerating at $z\approx0.7$. 

Especially given the profound mystery of dark energy, we should not 
fail to explore it even during the time that it amounts to only a few  
percent of the total energy density (i.e.\ about as much as baryons 
are today). Moreover, the manner in which dark energy rises could 
provide important insights into its nature. Consider going all the way 
back to the epoch of cosmic microwave background (CMB) last scattering: at 
$z\approx1090$, the energy density fraction due to a cosmological 
constant was only about $10^{-9}$ of the total, however in other dark 
energy models the contribution could be $10^{-3}$ -- six orders of 
magnitude difference! (Current constraints on the early dark energy 
fraction are somewhat model dependent, but generally below $10^{-2}$ 
\cite{earlyde}.) We do not have clear ideas for how to detect 
contributions in this range, but next generation galaxy surveys will  
probe $z\gtrsim2$. Though the differences between models are less there, 
the absolute level of dark energy density is higher. 

Consider models that may define the envelope of viable, smoothly evolving 
dark energy models. Our fiducial cosmology is \lcdm, where the dark 
energy is a cosmological constant, $w=-1$. One may come up with a wide 
variety of behaviors for other models, but they must be viable under 
current constraints. In particular, they must nearly preserve the distance 
to CMB last scattering, accurately measured by CMB surveys. This involves 
a redshift integral over the dark energy density, so one could imagine 
models where a dramatic departure at one redshift is compensated by an 
opposite departure at another redshift, outside the range where other 
probes can see. These are somewhat constrained by the integrated 
Sachs-Wolfe effect and the growth of matter structure (too long or too 
little time in full matter domination affects gravitational potentials and 
structure growth). We therefore assume monotonic dark energy evolution, 
i.e.\ a steady rise of dark energy influence. 

Dark energy models specifically designed to preserve the distance to 
CMB last scattering are known as mirage dark energy \cite{mirage}. 
These will form an envelope around \lcdm, i.e.\ the viable models with 
predictions furthest from \lcdm\ and most difficult to detect. 
Mirage dark energy has an equation of state $w(a)=w_0+w_a(1-a)$ 
with $w_a=-3.6(1+w_0)$, where $a=1/(1+z)$. Thus it is a one parameter 
family (with \lcdm\ a subcase with $w_0=-1$). We take our envelope of 
models to be those with $w_0=[-1.2,-0.8]$, hence between 
$(w_0,w_a)=(-1.2,+0.72)$ and $(-0.8,-0.72)$, with \lcdm\ in the middle. 
More extreme models are disfavored by lower redshift observations such 
as supernova and baryon acoustic oscillation (BAO) distances. 

\begin{figure}[!htb]
\centering 
\includegraphics[width=\columnwidth]{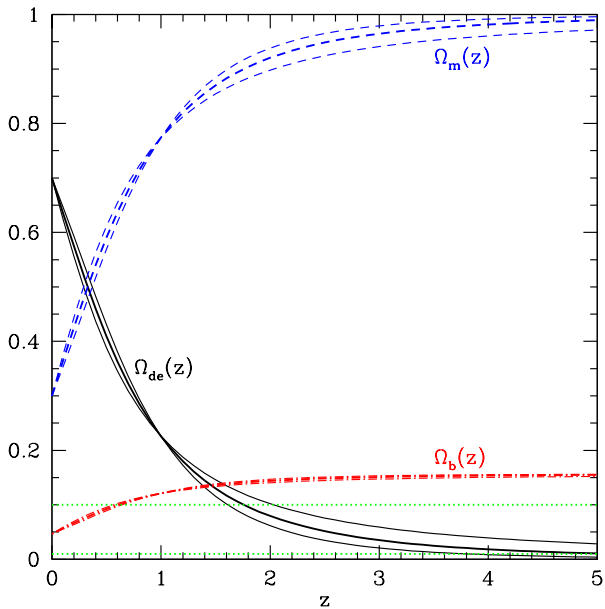} 
\caption{
The fractional energy densities in dark energy $\ode(z)$, all matter 
$\om(z)$, and baryons $\Omega_b(z)$ are shown vs redshift. The 
heavier middle curves are for \lcdm; the lighter curves are for mirage 
dark energy, with the $w_0=-1.2$ case lying above the $w_0=-0.8$ case 
at high redshift for $\ode(z)$ and below for $\om(z)$, $\Omega_b(z)$. 
The green dotted horizontal lines indicates 10\% and 1\% of the total 
energy density. 
} 
\label{fig:odezmir} 
\end{figure}

Figure~\ref{fig:odezmir} shows the dark energy density fraction  
of the total energy density as a function of redshift, $\ode(z)$, 
for \lcdm\ and the two envelope mirage dark energy models. 
At redshift $z=1.5$ we have $\ode(z)=0.146$, 0.130, 0.116 for the 
mirage $w_0=-1.2$, \lcdm, mirage $w_0=-0.8$ cases respectively; 
at $z=2$, $\ode(z)=0.102$, 0.080, 0.062; at $z=2.5$, $\ode(z)=0.076$, 
0.052, 0.035. (Throughout this article we take a flat cosmology with 
fiducial present matter density $\om=0.3$.) 
Thus at $z=1.5$ they all contribute over 10\%, and at 
$z=2.5$ the dark energy density is still several percent.

%%%%%%%%%%%%%%%%%%%%% 

%%%%%%%%%%%%%%%%%%  
\section{Constraining Dark Energy Density} \label{sec:method} 

Next generation cosmic surveys will use various methods to constrain 
cosmology; the leading probes for mapping evolution of a cosmic 
quantity such as $\ode(z)$ involve distances $d(z)$ such as derived from 
supernovae and BAO, radial distance intervals that can be viewed 
as a measure of the inverse Hubble parameter $H^{-1}(z)$ from 
radial BAO, and the growth rate of cosmic structure, $\fsig(z)$. 

\begin{figure}[!htb]
\centering 
\includegraphics[width=\columnwidth]{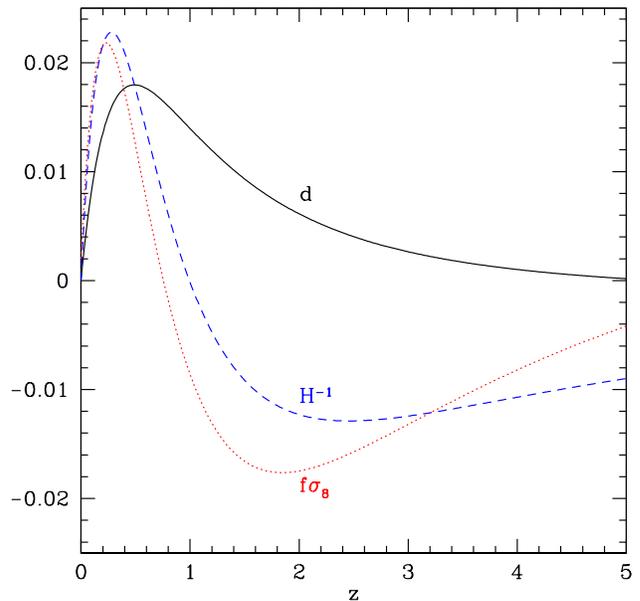} 
\caption{
The deviations $\Delta_X(z)$ between mirage dark energy with $w_0=-1.2$ 
and \lcdm \ are shown for $X=d$, $H^{-1}$, and $\fsig$. } 
\label{fig:dhmirdev} 
\end{figure}

The impact of the different rises of dark energy on these observables 
is presented in Figure~\ref{fig:dhmirdev} as the deviation 
\be 
\Delta_X(z)=\frac{X_{\rm mir12}(z)-X_\Lambda(z)}{X_\Lambda(z)}\,, 
\ee 
between mirage dark energy with $w_0=-1.2$ and \lcdm, 
for $X=d$, $H^{-1}$, and $\fsig$. (Mirage dark energy with $w_0=-0.8$  
looks similar, with the signs of the deviations flipped.)

%%%%%%%%%%%%%%%%%%%%% 

We see that observations need to be at the percent level to 
distinguish clearly differences in the rise of dark energy. As 
well known, the sensitivity to dark energy density is greatest 
at moderate redshifts $z<1$, and this is the target of the 
current generation of experiments. However, there is 
secondary sensitivity around $z\approx2$, and this is what we 
focus on in this article. At high redshift $z>3$ there is diminishing 
sensitivity to dark energy density. Our emphasis here is on 
the range $z=[1.5,2.5]$, so as to supplement current experiments 
and yet still have sensitivity to measure dark energy density. 
From Figure~\ref{fig:odezmir} we saw that at $z=2.5$, the 
\lcdm\ dark energy density fraction is 0.052 so a 2--$3\sigma$ 
detection of any dark energy density requires observational 
precision such that $\sigma(\ode(z=2.5))\approx0.017$--0.026. That 
would also give a $\sim1\sigma$ distinction between \lcdm\ 
and either mirage model. Weaker constraints would not be 
informative about the rise of dark energy. 

We emphasize that we adopted the mirage models simply 
as a reasonable envelope that allows some quantitative 
estimation of the magnitude of the impact from changing dark 
energy density. Going forward we instead employ a model  
independent approach. We allow the dark energy density 
to float freely in five independent redshift slices of width 
$dz=0.2$, from $z=[1.5,1.7]$ to $z=[2.3,2.5]$. 

The dark energy density in each slice, $\ode(z_i)$, where  $i=1-5$ 
is the slice number, is a new cosmological parameter that 
we constrain through the information matrix formalism. 
Transitions between slices are tanh smoothed, with a width  
of 0.01 in $\ln(1+z)$. In addition the present matter density 
$\om\equiv\om(z=0)$ is a free parameter, as is 
$\sigma_8\equiv\sigma_8(z=0)$ 
when we use $\fsig(z)$ as a probe. Outside the redshift slices 
the cosmology is fixed to the fiducial \lcdm. Thus our free 
parameters are 
$p=\{\om,\ode(z_1),\ode(z_2),\ode(z_3),\ode(z_4),\ode(z_5),\sigma_8\}$ 
and we apply the information matrix formalism to constrain them. 

That is, 
\be 
F^X_{jk}=\sum_{z_i} \frac{\partial X(z_i)}{\partial p_j}\,\frac{\partial X(z_i)}{\partial p_k}\,\frac{1}{\sigma^2(X_i)}\,,  
\ee 
where $X=d$, $H^{-1}$, $\fsig$ are the measurements, and 
$X_i\equiv X(z_i)$. These 
are of course not directly measurable, but derived from observations of 
supernova magnitudes and the galaxy power spectrum BAO and redshift 
space distortions. However we seek a high level answer to the question 
of whether there is a dark energy density science case for observations 
at $z\gtrsim2$, before considerations of any survey specifics. 
Therefore we assume as a first step here that we can obtain 
measurements of precision $\sigma(X_i)$ in each redshift slice. We 
chose the slice widths $dz=0.2$ so that the correlations between 
measurements in different slices should be ignorable (of course real 
world surveys need to ensure that common mode systematics are 
small after correction). In addition, the measurements via different probes 
are taken to be independent, so the measurement noise matrix is 
diagonal, simply  $\sigma^2(X_i)$. Again, to go beyond this first 
estimation of constraints would require detailed survey design. 

We now have all the elements necessary for parameter constraints, 
with $\sigma(p_j)=\sqrt{(F^{-1})_{jj}}$, marginalized over the other parameters. 
We consider probes singly and in combination. Since next generation 
experiments will also have external constraints from current generation 
experiments, we also study the effect of Gaussian priors $\pi(\om)=0.01$ 
and $\pi(\sigma_8)=0.02$, and variations. For the probe precisions, 
we take these to be fractionally constant for all five redshift slices, i.e.\ 
$\sigma(X_i)=\sigma_{X,0}\,X_i$. For next generation surveys we take the 
precisions $\sigma_{X,0}$ 
on $d$, $H^{-1}$, $\fsig$ to be 1\%, 1.4\%, 1\% respectively. 
For radial BAO, taking its precision to be $\sqrt{2}$ times the 
transverse BAO precision is reasonable, given its two times fewer 
modes. We also study variations of all these precisions.

%%%%%%%%%%%%%%%%%%%%%%%% 
\section{Results} \label{sec:results} 

Beginning with one probe at a time, we find that data in only the 
range $z=[1.5,2.5]$ has a difficult time constraining $\ode(z)$; 
the probe evolution is too slight over this limited redshift range 
and becomes degenerate with either a neighboring slice or the 
present matter density $\om$. However, once we apply either 
a prior on $\om$ (e.g.\ standing in for current generation experiments) 
or combine two probes, then the densities become well estimated. 

For $d(z)$ plus the $\om$ prior, the constraints on the dark energy 
densities are unexciting: $\sigma(\ode(z_i))\sim 0.3-0.6$. Even fixing 
$\om$ does not appreciably approve on these. The reason is that 
$d(z)$ is an integral measure and a change in $\ode(z_i)$ is diluted 
by the long path length out to $z\approx2$. We saw this as well in 
Fig.~\ref{fig:dhmirdev} where the deviation between the mirage model 
and \lcdm\ was  only  $\sim0.6\%$ at $z=2$, falling to 0.4\% at $z=2.5$. 

For $H^{-1}(z)$ plus the $\om$ prior, the situation is much better. Now 
the constraints are  $\sigma(\ode(z_i))\approx 0.027$ for each redshift 
slice. Furthermore, the slices are substantially uncorrelated, with 
correlation coefficients  $r\approx0.5$. This leverage is since 
$H^{-1}(z)$ is a point measurement, depending on conditions at that 
particular redshift and not diluted by others. 

For $\fsig(z)$ we must add a prior on the new parameter 
$\sigma_8\equiv\sigma_8(z=0)$ otherwise it is highly covariant with 
the dark energy density variations. We take external data, as from 
current generation experiments, to impose the prior $\pi(\sigma_8)=0.02$. 
Even so, this delivers only a weak $\sigma(\ode(z_i))\sim 0.3-0.4$,  again 
because $\fsig(z)$ is an integral measure (this time the growth from high 
redshift down to $z$). In fact, even fixing $\om$ does not alleviate this, 
nor does tightening the prior to $\pi(\sigma_8)=0.01$. 

Combining probes helps considerably however. Using both $d(z)$ and 
$H^{-1}(z)$ delivers $\sigma(\ode(z_i))\approx 0.023$, and further imposing 
the $\om$ prior reduces the uncertainty to 0.021. Without the prior, 
$\om$ is determined to $0.0065$, and the correlation coefficients between 
dark energy density slices is a modest $r\approx0.4$. The combination of 
$d(z)$ and $\fsig(z)$ does not do well, with $\sigma(\ode(z_i))\sim 0.25$,  
or 0.23 with $\om$ prior. The integral natures of $d(z)$ and $\fsig(z)$ do 
not allow incisive constraints. 
Using $H^{-1}(z)$ and $\fsig(z)$ enables $\sigma(\ode(z_i))\approx 0.021$,  
or 0.020 with $\om$ prior, with the former having $r\approx0.4$ and 
$\sigma(\om)=0.0088$. 
Finally, all three probes together yield $\sigma(\ode(z_i))\approx 0.0019$, 
with $r\approx0.2$ and $\sigma(\om)=0.0051$, so there is no need for 
a $\om$ prior. 

\begin{figure}[!htb]
\centering 
\includegraphics[width=\columnwidth]{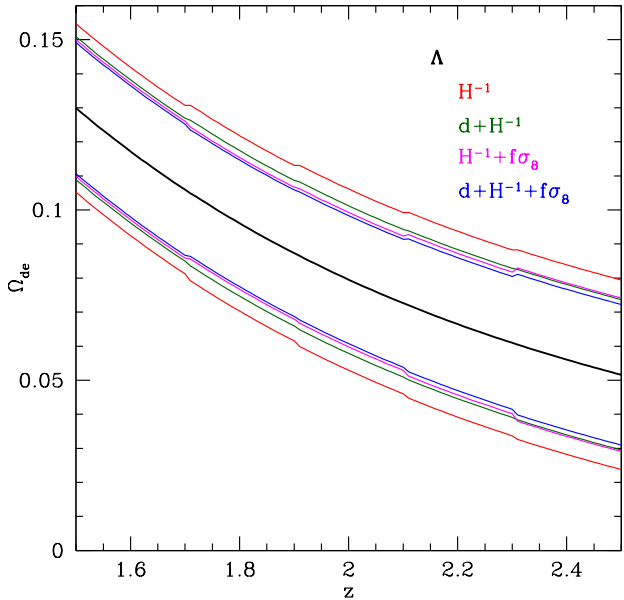} 
\caption{
Constraints on the dark energy density $\sigma(\ode(z_i))$ are shown 
for various combinations of probes, as $1\sigma$ bands around the 
fiducial \lcdm\ value (thick black curve). The probes used are $H^{-1}$ 
alone (red), $d+H^{-1}$ (dark green), $H^{-1}+\fsig$ (magenta), and 
$d+H^{-1}+\fsig$ (blue), from top to bottom at $z=2$. All combinations 
include a prior  $\pi(\om)=0.01$ although this is only really needed for 
the case of $H^{-1}$ alone. 
} 
\label{fig:bands} 
\end{figure}

Figure~\ref{fig:bands} shows the key results. We can see that $H^{-1}(z)$ 
(plus a $\om$ prior) provides the key leverage. Either $d(z)$ or $\fsig(z)$ 
can substitute for the $\om$ prior, and further tightens the constraint by 
a factor 1.2--1.3. All three together give a factor 1.4 improvement.

%%%%%%%%%%%%%%%%%%%%% 

A constraint on $\ode(z)$ of 0.02 would deliver a 6.5, 4.0, and  
$2.6\sigma$ evidence of nonzero dark energy density at $z=1.5$, 2, 
2.5 respectively, within a  \lcdm\ cosmology. The envelope spanning 
the mirage models would lie at $\approx\pm1\sigma$. Thus, while it 
would not in itself\footnote{Fortunately 
we also have the current generation of experiments probing $z<1.5$ 
in detail. These could, for example, narrow the deviation of $w_0$ from 
$-1$ and hence constrain mirage models.}  
distinguish between reasonable, viable monotonically 
evolving dark energy models, if the results lay outside the envelope 
(and hence greater than $1\sigma$ from \lcdm) then we would have 
not simply a 
hint of dark energy different from \lcdm, but dark energy 
behaving in a particularly exotic manner. 

We can examine the scaling of the results with the measurement 
precisions, and the prior, to see how the constraints improve or 
degrade with stronger or weaker experiments. With all three probes 
combined, having no prior on  $\om$ vs fixing it changes 
$\sigma(\ode(z_i))$ from 0.020 to 0.018. Each of the two probe 
combinations shows a similar 10\% change in constraints. Only for 
$H^{-1}(z)$ alone is the level of the $\om$ prior influential, with 
fixing $\om$ tightening the density constraints to 0.019 from 0.027. 
Conversely, loosening the prior to $\pi(\om)=0.02$ weakens this 
constraint to 0.042 (but has no effect on the two and three probe 
combinations). 

If the precision on $H^{-1}$ measurements weakens by a factor 2, 
to 2.8\%, then the density constraints from $H^{-1}$ plus the $\om$  
prior weaken by a factor 1.5, to $\sigma(\ode(z_i))\approx0.041$. 
With both $d$ and $H^{-1}$ precisions worse by a factor 2, then 
$\sigma(\ode(z_i))\approx0.040$. For all three precisions, including 
$\fsig$, worse by a factor 2, then $\sigma(\ode(z_i))\approx0.037$. 
Conversely, if $H^{-1}$ is better by a factor 2 than our baseline 
(hence 0.7\%), then 
with the $\om$ prior we find $\sigma(\ode(z_i))\approx0.021$. For 
$d+H^{-1}$  better by a factor 2, then $\sigma(\ode(z_i))\approx0.011$. 
Adding also $\fsig$ still at 1\% yields $\sigma(\ode(z_i))\approx0.011$ too. 
For such precision on $H^{-1}$, when combined with $d$ the 0.01 prior 
on $\om$ is irrelevant. Density estimation uncertainties at the 0.011 
level would deliver $>5\sigma$ evidence for dark energy in each redshift 
slice, and distinguish \lcdm\ from the mirage dark energy envelope at 
$\sim2\sigma$.

%%%%%%%%%%%%%%%%%%%%%%%%% 
\section{Conclusion} \label{sec:concl} 

The epoch before dark energy domination was not one of pure matter 
domination. Rather, the dark energy density is still expected to constitute 
$>5\%$ of the total energy density (for comparison, greater than the 
baryon energy density today) out to $z=2.5$. The precisions required to 
make an informative measurement of the dark energy density at such 
redshifts -- first, detection of a nonzero contribution, and second, 
measurement of the value for comparison  between \lcdm\ and other 
dark energy models -- are just outside the reach of the currently starting 
generation of dark energy experiments. 

We demonstrate that next generation experiments with precision 
measurements of transverse and radial  BAO, and the growth rate 
from redshift space distortions, over the range $z=1.5-2.5$, can achieve 
significant constraints on dark energy in this epoch. We set the scale of 
constraints to aim for by considering the envelope of smoothly evolving 
dark energy models that preserve the distance to CMB last scattering 
(``mirage dark energy''), but then carry out the detailed analysis in a 
model independent manner through information matrix constraints on 
independently varying dark energy densities in redshift slices. 

Measurements of $H^{-1}(z)$ through radial BAO are the most incisive, 
due to their local, nonintegral, nature. Combining this with distances 
$d(z)$ from transverse BAO, or $\fsig(z)$ from redshift space distortions, 
delivers constraints at the level of $\sigma(\ode(z_i))\sim0.02$. This 
corresponds to detection of dark energy at the level of $\sim6.5\sigma$ 
at $z=1.5$ to $\sim2.6\sigma$ at $z=2.5$. Distinction between \lcdm\ 
and mirage dark energy will require combination with the current, lower 
redshift experiments. 

Surveys extending deeper than $z=2.5$ run into the 
double issue of increasing difficulty to achieve the necessary 
measurement precision while also facing declining probe sensitivity 
to dark energy density at the higher 
redshifts\footnote{In 
the final stages of this work, we learned that 
surveys with a wide range of redshifts $z>2$ 
are explored in \cite{sailer}.}. 
Examining the impact of measurement precision over $z=1.5-2.5$, 
we confirm that $H^{-1}$ has the greatest leverage, and a survey  
strengthening its accuracy to the 0.7\% level could deliver 
$\sigma(\ode(z_i))\approx0.011$. 
This could give significant insight on the rise of dark energy in our 
universe.

%%%%%%%%%%%%%%%%% 
\acknowledgments 

This work is supported in part by the Energetic Cosmos Laboratory and by the 
U.S.\ Department of Energy, Office of Science, Office of High Energy 
Physics, under contract no.~DE-AC02-05CH11231.

%%%%%%%%%%%%%%%%%%%%%%% 

\end{document}